\documentclass[fleqn,usenatbib]{mnras}

\usepackage[T1]{fontenc}

\DeclareRobustCommand{\VAN}[3]{#2}
\let\VANthebibliography\thebibliography
\def\thebibliography{\DeclareRobustCommand{\VAN}[3]{##3}\VANthebibliography}

\usepackage{graphicx}	
\usepackage{amsmath}	
\usepackage{amssymb}	
\usepackage{booktabs}   
\usepackage{multicol}
\usepackage{multirow}

\usepackage{newtxtext,newtxmath}

\title[HDE~228766 contains a Wolf$-$Rayet star?]{Using a new spectral disentangling approach to ascertain whether the massive binary HDE~228766 contains a Wolf$-$Rayet star\thanks{Based on observations collected at Observatorio Astron\'omico Nacional de San Pedro M\'artir (M\'exico) and Observatoire de Haute Provence (France).}}

\author[E. A. Quintero et al.]{
Edwin A. Quintero,$^{1}$\thanks{E-mail: equintero@utp.edu.co}
and Philippe Eenens,$^{2}$
\\
$^{1}$Observatorio Astronómico (OAUTP), Universidad Tecnológica de Pereira, Carrera 27 \#10-02, 660003 Pereira, Colombia\\
$^{2}$Departamento de Astronomía, Universidad de Guanajuato, Callejón de Jalisco S/N, 36000 Guanajuato, México.\\
}

\date{Accepted XXX. Received YYY; in original form ZZZ}

\pubyear{2023}

\begin{document}
\label{firstpage}
\pagerange{\pageref{firstpage}--\pageref{lastpage}}
\maketitle

\begin{abstract}
The massive binary HDE~228766 is composed of an O type primary and an evolved secondary. However, previous qualitative analyses of the composite spectrum have led to a wide discussion about whether the secondary is an Of or a Wolf$-$Rayet star. We use new observations and our novel QER20 package to disentangle for the first time the spectra of the two stellar components and obtain artifact$-$free reconstructed spectra, yielding the more accurate and reliable spectral classifications of O7.5 V((f))z for the primary and O6 Iaf for the secondary. The emission features of the P$-$Cygni profiles of the H\,$\beta$ and He~{\sc i} 5876 \AA\ lines, present in the reconstructed spectrum of the secondary, show that this star is at an initial phase of its transition to the WN evolutionary stage. A previously unobserved variable emission, composed of at least four independent features, is seen since 2014 superposed to the H\,$\gamma$ absorption line. Our analysis reveals that these emission features originate from a physically extended region. This could be explained by an episode of enhanced mass loss in the scenario of a non$-$conservative evolution of the binary.
\end{abstract}

\begin{keywords}
binaries: spectroscopic -- stars: massive -- stars: winds -- stars: individual (HDE~228766)
\end{keywords}



\section{Introduction}\label{sec:1_intro}

HDE~228766, a massive binary system with a period of 10.7426 days, is of special interest because it is very likely that it has undergone episodes of mass transfer and that its secondary component is now in transition toward the Wolf$-$Rayet (WR) phase.~\cite{Hiltner1951} classified the secondary star of this system as WN, attributed to this component the emission features present in the observed spectra, and surmised that the absorption profiles are due to a B or late O primary. Additionally,~\cite{Hiltner1951} attributed all the absorption of the H\,$\gamma$ line to the late O component, and observed that the orbital modulation of this line is more irregular and of smaller amplitude compared to the other lines present in the observed spectra. This led Hiltner to conclude that the primary should be much more massive than the secondary. 

\cite{Hiltner1966} and~\cite{Walborn1973b} classified the secondary component of HDE~228766 as type Of. This classification implies that this star contributes to the absorption flux of the H\,$\gamma$ line in the observed spectra. Therefore,~\cite{Walborn1973b} recommended reevaluating the orbital parameters and in particular the high mass ratio established by~\cite{Hiltner1951} for this system. Likewise,~\cite{Massey1977} suggested that the secondary has its own absorption spectrum, which led them to classify this system as O\,7.5 + O\,5.5f. However, for~\cite{Massey1977} the presence of emission lines in the spectrum of HDE~228766 indicates that the secondary is already evolving toward the WR phase, a process that could take 500,000 years before reaching a typical WR mass of ($\approx$10 $M_ \odot$) at a mass$-$loss rate of 10$^{-5}$ $M_ \odot$ yr$^{-1}$. 

In contrast,~\cite{Rauw2002a} classified the secondary of HDE~228766 as WN8ha, based on an estimated $EW$ of $\approx$ 10\AA\ for the He~{\sc ii} 4686 \AA\ emission line, and a comparison of the observed spectra with those of WR 25 and WR 108. Yet, comparison with these objects could be biased since~\cite{Gamen2006} reported that WR 25 is binary, while~\cite{VanderHucht2001} and~\cite{Shara2009} mention that WR 108 is a probable WN + OB binary system. It should be noted that previous spectral classifications were not quantitative and were not based on disentangled spectra. 

On the other hand,~\cite{Rauw2002a} discussed the interaction between the winds of the components of HDE~228766 from the identification of a weak variable emission feature in the H\,$\alpha$ line. Analysis of this spectral feature led~\cite{Rauw2002a} to conclude that the interaction between the winds occurs in a region located near the surface of the primary star. From an X$-$ray spectroscopic study,~\cite{Rauw2014a} concluded that the wind collision zone is between 0.67 $d$ and 0.71 $d$ of the secondary component, where $d$ is the orbital separation between the stars.

In this paper we aim at clarifying the evolutionary status of the secondary through the analysis of disentangled spectra free of artefacts. We use newly available observations of HDE~228766 to update the orbital parameters calculated for this system~(\S~\ref{sec:2_obs}), and to perform for the first time the spectral disentangling of the observed spectra applying the novel QER20 Package algorithm (\S~\ref{sec:3_separacion}). Based on the reconstructed spectra, we propose a quantitative spectral classification and discuss the evolutionary status of the stellar components~(\S~\ref{sec:4_clasificacion}). Finally, we analyze the variable emission feature that we discovered in the H\,$\gamma$ line and discuss the location of the emitting zone~(\S~\ref{sec:5_vientos}).

\section{Observed spectra and orbital parameters of HDE~228766}\label{sec:2_obs}

In this study we use 75 spectra of HDE~228766 collected between 1999 and 2021 at the Observatoire de Haute Provence (OHP), France, and the Observatorio Astronómico Nacional of San Pedro Mártir (SPM), M\'exico. Table~\ref{tab:obs} lists the spectral range, the spectral resolution and the number of observations (\textit{N}) collected during each observing campaign.

The observations collected at the OHP, previously used in the~\cite{Rauw2002a} study, were obtained with the 1.52 m telescope equipped with the Aurélie spectrograph. In 1999, the detector was a Thomson TH7832 linear array with a pixel size of 13 $\mu$m$^{2}$. From 2000 on, the detector of the Aurélie instrument was replaced by a 2048 × 1024 CCD EEV 42-20\#3, with a pixel size of 13.5 $\mu$m$^{2}$. We refer to~\cite{Rauw2002a} for an explanation of the procedure followed to reduce these data. 

The SPM spectra were collected with the ESPRESSO spectrograph mounted on the 2.12 m telescope. The CCD detector was a E2V optical chip with 2048 × 2048 pixels of 13.5 $\mu$m$^{2}$. These spectra were reduced using the \texttt{MIDAS} echelle package. To remove the telluric lines we use the \texttt{telluric} task of \texttt{IRAF} and the spectral atlas of~\cite{Hinkle2000}.

\begin{center}
\begin{table}
\centering
\caption{Observations used in this study of HDE~228766. The last column indicates the number of observations collected during each campaign (\textit{N}), in the given spectral range, at the respective observatory (Obs.).\label{tab:obs}}%
\tabcolsep=0pt%
\begin{tabular*}{20pc}{@{\extracolsep\fill}lcccr@{\extracolsep\fill}}
\toprule
\textbf{Campaign} & \textbf{Obs.}  & \textbf{\begin{tabular}[c]{@{}c@{}}Spectral Range\\ \AA\end{tabular}}  & \textbf{\begin{tabular}[c]{@{}c@{}}Resolution\\ \AA\end{tabular}}  & \textbf{\textit{N}} \\
\midrule
Jul - Aug 1999    & OHP     & 4150 - 4920       & 0.05      & 20         \\
Sep - Oct 2000    & OHP     & 4450 - 4900       & 0.10      & 10         \\
Sep - Oct 2000    & OHP     & 6350 - 6769       & 0.12      & 9          \\
Sep - Oct 2001    & OHP     & 6350 - 6769       & 0.12      & 10         \\
Jun 2014          & OHP     & 4450 - 4900       & 0.10      & 6          \\
Jun 2014          & SPM     & 4150 - 7260       & 0.12      & 11         \\
Aug 2015          & SPM     & 4200 - 7270       & 0.12      & 3          \\
Sep 2017          & OHP     & 4450 - 4900       & 0.10      & 5          \\
Sep 2021          & SPM     & 3840 - 7500       & 0.12      & 1          \\ 
\bottomrule
\end{tabular*}
\end{table}
\end{center}

\subsection{The composite spectrum of HDE~228766}\label{subsec:2.1_spec}

\begin{figure*}
\centerline{\includegraphics[width=2.5\columnwidth]{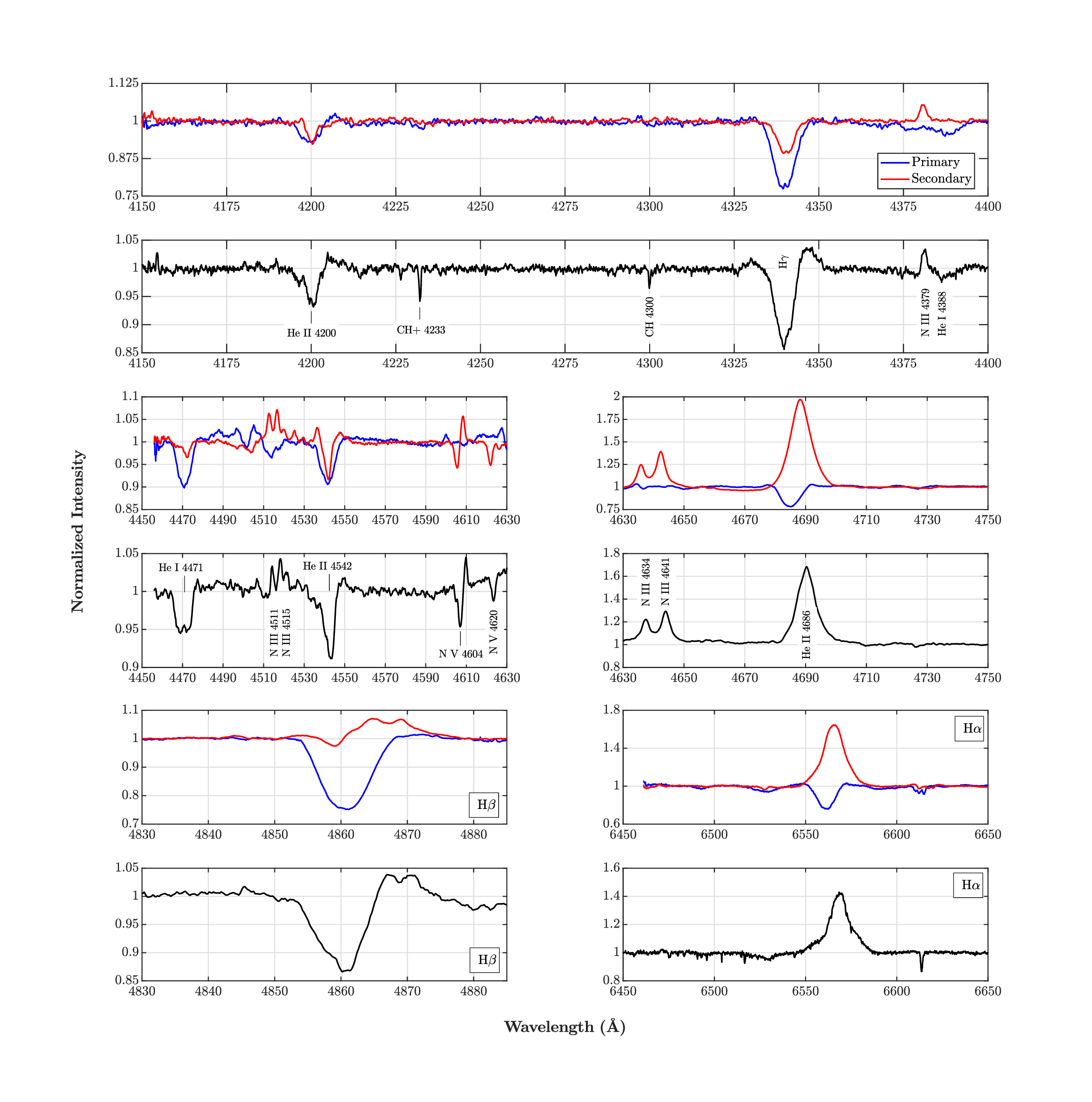}}
\caption{Reconstructed spectra of the primary (blue) and secondary (red) components of HDE~228766, obtained using the QER20 Package algorithm~\citep{Quintero2020}. The spectra are scaled assuming the light ratios 43.1\% and 56.9\% for the primary and secondary, respectively (see Section~\S~\ref{sec:3_separacion} for details). For comparison, one of the spectra of HDE~228766 collected at the Observatorio Astronómico Nacional San Pedro Mártir during the June 2014 campaign is shown in black (phase $\phi = 0.14$). The emission wings around the H\,$\gamma$ line are variable and will be discussed in Section~\S~\ref{sec:5_vientos}. The region around the He~{\sc i} 5876 \AA\ line is displayed in Figure~\ref{fig:5876}.\label{fig:reconstruidos}}
\end{figure*}

The composite spectrum of HDE~228766 contains the typical emissions of Of and WN stars at N~{\sc iii} 4634$-$41 \AA\ and He~{\sc ii} 4686 \AA\ (Figure~\ref{fig:reconstruidos}, in black). The H\,$\alpha$ line is also present in strong emission. The lines He~{\sc ii} 4200 \AA, He~{\sc ii} 4542 \AA\ and N~{\sc v} 4604 \AA\ exhibit a relatively weak P$-$Cygni profile associated with the secondary wind, while in the H\,$\beta$ line this profile is much stronger. The clear splitting of the He~{\sc i} 4471 \AA\ line by the Doppler effect demonstrates the binarity of the system. The observed spectra contain absorption lines of the interstellar medium due to CH$^+$ at 4233\,\AA\ and CH at 4300\,\AA.    

\begin{figure}	         
\centerline{\includegraphics[width=\columnwidth]{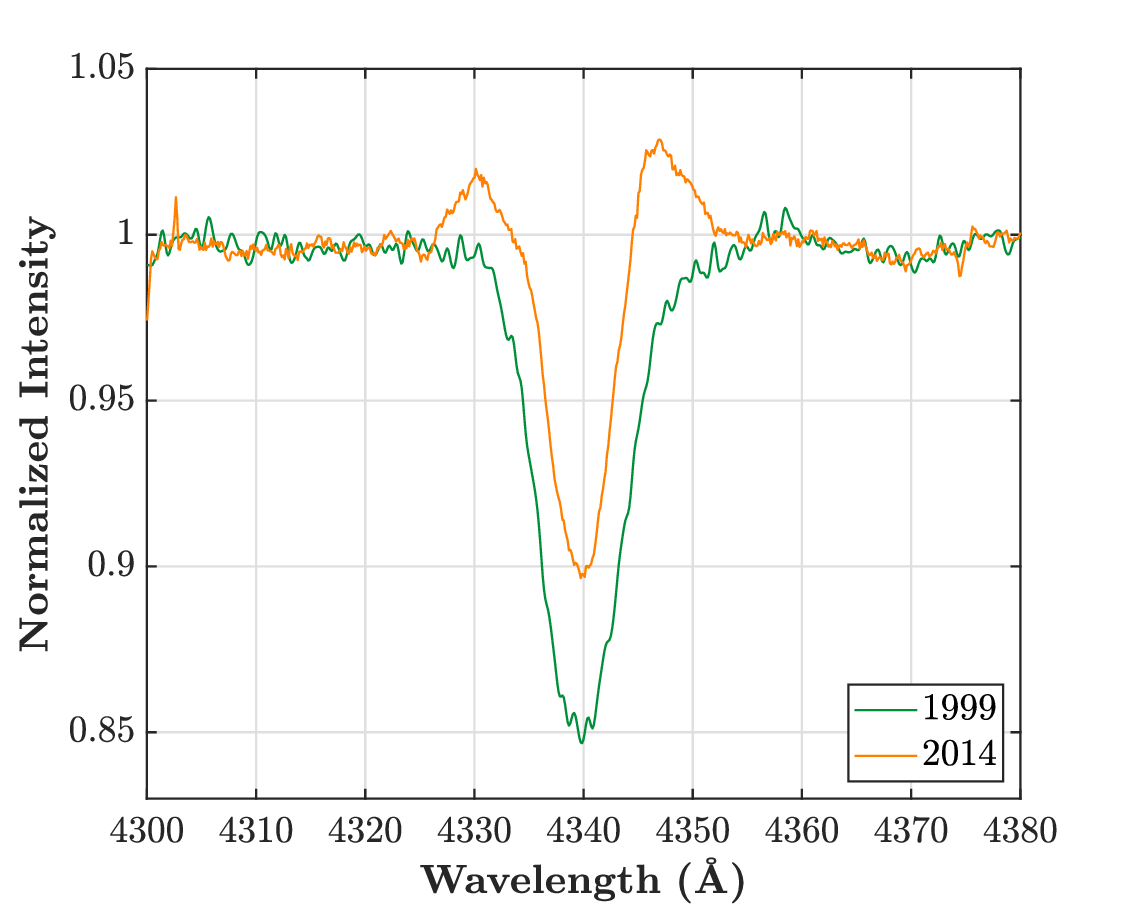}}
	\caption{Averages of observations of HDE~228766 collected during the 1999 (green) and 2014 (orange) campaigns, centered on the H\,$\gamma$ line.\label{fig:1999vs2014}}
\end{figure}

In the 20 observations from the 1999 campaign that include the H\,$\gamma$ line, this line appears entirely in absorption over the entire orbital cycle. In contrast, emission wings located on both sides of the H\,$\gamma$ absorption are observed in all our spectra between 2014 and 2021. Figure~\ref{fig:1999vs2014} shows the H\,$\gamma$ line averaged over the 1999 campaign (green) and over the 2014 campaign (orange). 

\begin{figure}	         
\centerline{\includegraphics[width=\columnwidth]{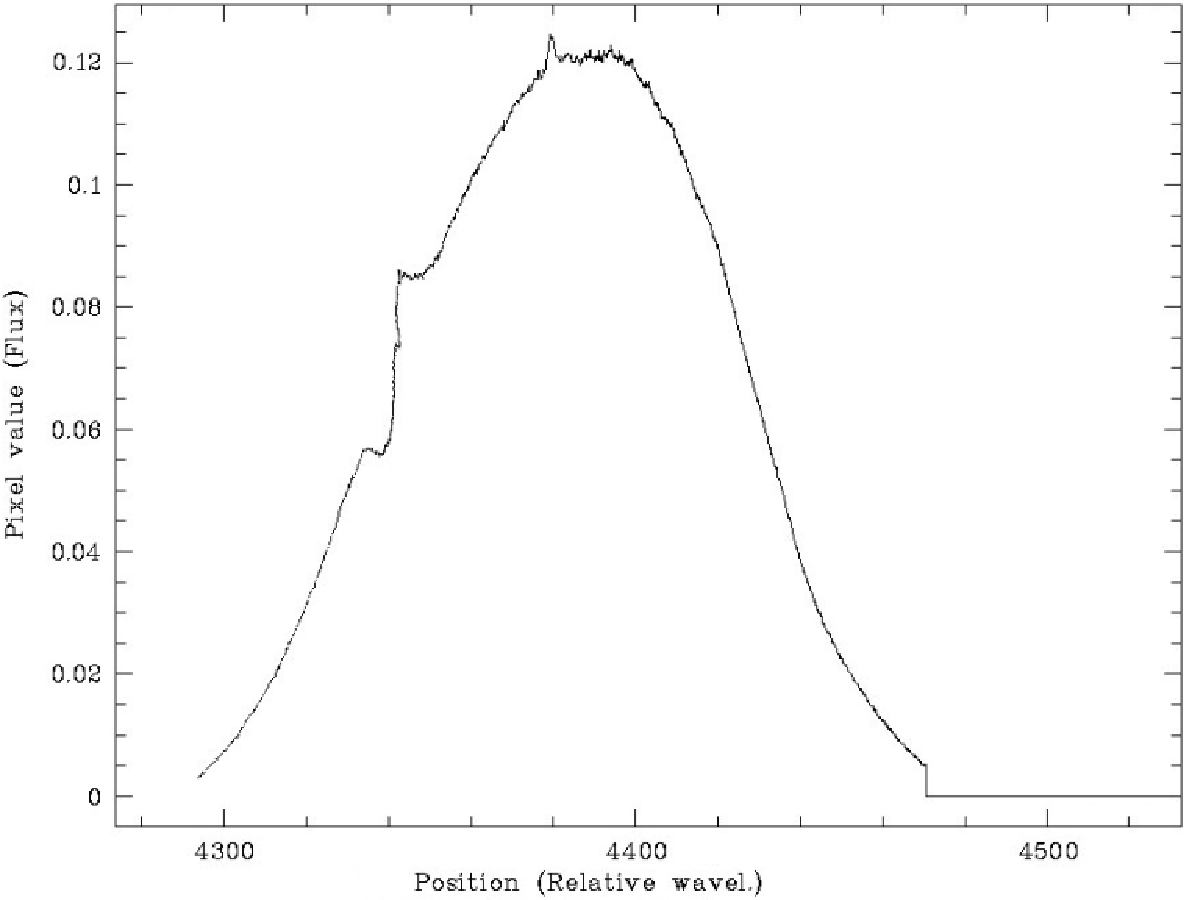}}
	\caption{Uncorrected Echelle order of one 2014 spectrum of HDE~228766, clearly showing  
	 emission wings on both sides of the H\,$\gamma$ line (clearly visible on the red side). These wings are neither
	the effect of the blaze correction nor of the normalization. \label{fig:Blaze}
	}
\end{figure}

Figure~\ref{fig:Blaze} shows the appearance 
of one of the spectra collected at SPM during the 2014 campaign, before correcting the blaze and fitting the continuum. In this initial step of the reduction of the spectra, the existence of the emission wings that we discovered in the H\,$\gamma$ line is evident, which allows us to infer that they are not an artefact of the reduction process.

In all our spectra from 2014 this 
excess emission is clearly visible at H\,$\gamma$, as it contrasts with the absorption profile observed previously. It is probable that there are excess emissions in the H\,$\alpha$ and/or H\,$\beta$ spectral lines, but these are not clearly identified in the observations due to the strong emissions from the secondary. In fact,~\cite{Rauw2002a} reported a weak emission feature in H\,$\alpha$ that we do not clearly identify in all observations (as does occur in the case of H\,$\gamma$). In \S~\ref{sec:5_vientos} we analyze this variable emission feature in H\,$\gamma$ line and discuss its possible interpretation.

\subsection{Orbital parameters of HDE~228766}\label{subsec:2.2_losp}

\begin{figure}
	\centerline{\includegraphics[width=\columnwidth]{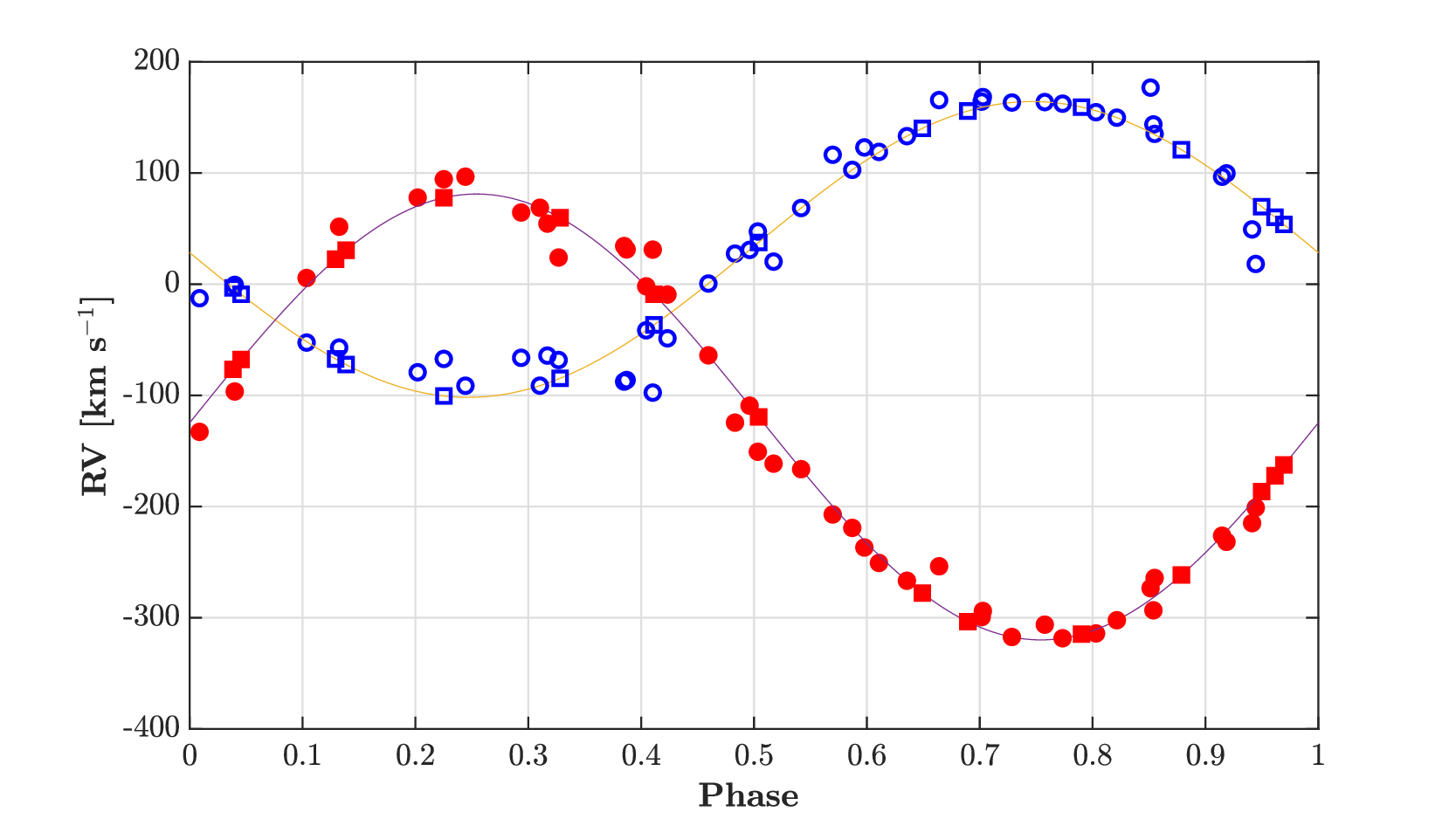}}
	\caption{Radial velocities of the primary (blue) and secondary (red) components of HDE~228766, derived from the Doppler shift of the He~{\sc ii} 4542 \AA\ line. Open symbols represent the radial velocities of the primary component, while the closed symbols correspond to the secondary. Circles indicate OHP observations, while squares correspond to SPM.\label{fig:RVs}}
\end{figure}

Figure~\ref{fig:RVs} illustrates the radial velocity for the primary (blue) and secondary (red) components, measured from the Doppler shift of the He~{\sc ii} 4542 \AA\ line. The set of collected observations adequately covers the phases of the orbital cycle, which is essential for the spectral disentangling (\S~\ref{sec:3_separacion}). Phase 0 corresponds to the conjunction phase, with the secondary component located between the primary star and the observer. 

We use the new observations from the 2014 -- 2021 (SPM) campaigns to update the orbital parameters of HDE~228766 calculated by~\cite{Rauw2002a}. The orbital solution was computed using the radial velocities measured on the He~{\sc ii} 4542 \AA\ line (Figure~\ref{fig:RVs}) and the \texttt{LOSP}\footnote{\texttt{LOSP} is available in: \url{https://www.stsci.edu/~hsana/losp.html}} program (Li\`ege Orbital Solution Package), in its version for the SB2$-$circular case~\citep{Rauw2000,Sana2006}, with the period fixed at 10.7426 days and eccentricity equal to zero~\citep{Rauw2002a}. We also compute the orbital solution using the \texttt{Binary Star Solve}\footnote{\texttt{Binary Star Solve} is available in: \url{https://github.com/NickMilsonPhysics/BinaryStarSolver}} program~\citep{Milson2020}, developed in \texttt{Python}. Table~\ref{tab:losp} compares the orbital parameters obtained with these two methods, as well as those presented in Table 3 of~\cite{Rauw2002a}. The results are consistent with each other. However, the error bars given by \texttt{Binary Star Solve} are much smaller than those obtained using \texttt{LOSP}. This is because, for the calculation of the errors, \texttt{LOSP} uses the classical error propagation theory, while \texttt{Binary Star Solve} uses the inverse of the Hessian matrix as the covariance matrix to calculate likelihood estimates.

\begin{center}
\begin{table*}
\centering
\caption{Orbital parameters of HDE~228766 calculated from the radial velocities measured on the He~{\sc ii} 4542 \AA\ line (Figure~\ref{fig:RVs}). The time of conjunction ($T_{0}$) indicates the time at which the secondary component is located between the primary star and the observer (phase $\phi = 0$).\label{tab:losp}}%
\footnotesize
\begin{tabular}{cccccccc} 
\cmidrule[\heavyrulewidth]{1-8}
\multirow{2}{*}{\begin{tabular}[c]{@{}c@{}}\textbf{Orbital}\\\textbf{Parameter}\end{tabular}} & \multirow{2}{*}{\textbf{Symbol}} & \multicolumn{2}{c}{\textbf{This Work $-$ LOSP}} & \multicolumn{2}{c}{\textbf{This Work $-$ Binary Star Solve}} & \multicolumn{2}{c}{\textbf{Rauw et al. (2002)}}  \\ 
\cline{3-8}
                &             & \textbf{Primary} & \textbf{Secondary}    & \textbf{Primary} & \textbf{Secondary}   & \textbf{Primary} & \textbf{Secondary}  \\ 
\midrule
\begin{tabular}[c]{@{}c@{}}Period (days)\end{tabular}   & $P$    & \multicolumn{2}{c}{10.7426 (fixed)}    & \multicolumn{2}{c}{10.7426 (fixed)}      & \multicolumn{2}{c}{10.7426 (fixed)}         \\
\midrule
\begin{tabular}[c]{@{}c@{}}Eccentricity\end{tabular}   & $\epsilon$    & \multicolumn{2}{c}{0 (adopted)}    & \multicolumn{2}{c}{0 (adopted)}      & \multicolumn{2}{c}{0 (adopted)}         \\
\midrule
\begin{tabular}[c]{@{}c@{}}Conjunction time\\(HJD $-$ 2 440 000)\end{tabular}   & $T_{0}$       & \multicolumn{2}{c}{19481.530 $\pm$ 0.02}    & \multicolumn{2}{c}{19481.530 $\pm$ 0.02}   & \multicolumn{2}{c}{1822.052 $\pm$ 0.305}        \\
\midrule
\begin{tabular}[c]{@{}c@{}}Systemic\\velocity (km\,s$^{-1}$)\end{tabular}   & $\gamma$      & 31.5 $\pm$ 4.0    & -119.2 $\pm$ 5.0  & 32.0 $\pm$ 0.2  & -119.8 $\pm$ 0.2   &   31.4 $\pm$ 7.3      &    -119.0 $\pm$ 8.2    \\
\midrule
\begin{tabular}[c]{@{}c@{}}Semi-amplitude of\\the radial velocity\\curve (km\,s$^{-1}$)\end{tabular}   & $K$    & 134.0 $\pm$ 4.9   &  200.5 $\pm$ 7.3 & 133.0 $\pm$ 0.3   &  201.0 $\pm$ 0.3    &    133.7 $\pm$ 7.6          &    200.2 $\pm$ 10.9       \\
\midrule
\begin{tabular}[c]{@{}c@{}}Semi-major axis\\($R_{\odot}$)\end{tabular}   &  $a$ $sin$ $i$   &  28.4 $\pm$ 1.0  &  42.6 $\pm$ 1.5              &-----    &   -----   &   28.4 $\pm$ 1.6     &   42.5 $\pm$ 2.3      \\
\midrule																	
\begin{tabular}[c]{@{}c@{}}Mass function\\($M_{\odot}$)\end{tabular}   &  $f(m)$   &  2.7 $\pm$ 0.3  &  9.0 $\pm$ 1.0 
& 2.6 $\pm$ 0.1    &   -----     &   -----      &   -----    \\		
\midrule
\begin{tabular}[c]{@{}c@{}}Mass\\($M_{\odot}$)\end{tabular}   &  $m$ $sin^{3}$ $i$   &  25.0$\pm$ 2.1  &  16.7 $\pm$ 1.3 
& -----    &   -----       &  24.8 $\pm$ 1.8      &  16.6 $\pm$ 2.4    \\		
\midrule
\begin{tabular}[c]{@{}c@{}}Mass ratio\end{tabular}     & $q = M_{1}/M_{2}$   & \multicolumn{2}{c}{1.5 $\pm$ 0.1}    & \multicolumn{2}{c}{-----}      & \multicolumn{2}{c}{1.50 $\pm$ 0.11}            \\				
\midrule
\begin{tabular}[c]{@{}c@{}}Radius of the\\Roche lobe ($R_{\odot}$)\end{tabular}   &  $R_{RL}$ $sin$ $i$   &  29.4 $\pm$ 1.1  &  24.5 $\pm$ 0.9 
& -----    &   -----     &  29.3 $\pm$ 1.5        &  24.4 $\pm$ 1.2      \\							\bottomrule\end{tabular}
\end{table*}
\end{center}

Among the results of the orbital solution of this binary, the high mass ratio of the system and the high minimum masses of the stellar components stand out. The mass ratio could be explained by an episode of mass transfer from the secondary to the primary due to a Roche lobe overflow episode. On the other hand, the high minimum masses, added to the low orbital period of the system and the proximity between the stars, allowed~\cite{Rauw2002a} to infer the existence of a stellar wind shock region that could produce relatively intense optical and X$-$ray emission. In fact,~\cite{Rauw2014a} found that the X$-$ray spectra of this system exhibit variations that are consistent with what is expected for a wind$-$wind collision. Yet, the only evidence of this collision that~\cite{Rauw2014a} found in the optical spectra is reduced to a weak excess of variable emission in the H\,$\alpha$ line. This situation makes even more relevant our discovery of the relatively intense asymmetric variable emission wings presented by the H\,$\gamma$ line in the spectra of the 2014 -- 2021 campaigns, since it could be new evidence of the interaction between the stellar components of this system (\S~\ref{sec:5_vientos}).

\section{Spectral disentangling}\label{sec:3_separacion}

As seen in Figure~\ref{fig:reconstruidos}, the spectra of the stellar components of HDE~228766 are strongly blended, which greatly hinders spectral disentangling. Furthermore, by applying the widely used shift$-$and$-$add spectral disentangling technique proposed by~\cite{Marchenko1998}, huge artifacts caused by the wide and intense emission lines of the secondary component overshadow the relatively weak absorptions of the primary (this may explain why no attempts to disentangle the spectra of this system are found in the literature). These artifacts consist of broad spurious wings, located symmetrically with respect to the wavelength of the spectral lines. In the case of the primary component, these wings appear in absorption, while in the spectrum of the secondary they are observed in emission. \citet{Quintero2020} presents a detailed analysis of the shortcomings of the shift$-$and$-$add method. 

This difficulty in disentangling the spectra of HDE~228766 explains why until now the evolutionary status of the stars belonging to this system has not been established from a quantitative analysis of the individual spectra of the components. 

In~\cite{Quintero2020} we introduced a novel spectral disentangling method, known as the QER20 Package, ideal for binary systems with low Doppler shifts and wide and intense spectral lines. This package aims at correcting the artifacts produced by the shift$-$and$-$add technique and obtaining reconstructed spectra which are free of artifacts and thus yield more reliable line profiles and fluxes (see~\cite{Quintero2020} for a detailed description of the QER20 Package algorithm and the validation tests we performed using synthetic and real spectra). As described below, we have expanded the original version of the QER20 Package, which we successfully applied to the 9 Sgr binary system in~\cite{Quintero2020}, so that it can handle a wider variety of cases, given that each binary presents new challenges.

The fundamental principle of the QER20 Package consists in considering the integrated flux of a given spectral line as a free parameter. During each iteration, the QER20 Package looks for artifacts in the reconstructed profiles, and redistributes the flux of these artifacts between the reconstructed profiles of the two stars. However, the observed spectra of HDE~228766 proved to be a challenge for the QER20 Package, because the emission features of the P$-$Cygni profiles of the He~{\sc ii} 4200 \AA, He~{\sc ii} 4542 \AA, N~{\sc v} 4604 \AA\ and H\,$\beta$ lines can be treated as artifacts by the algorithm. In addition, the strong and wide emissions in the N~{\sc iii} 4634$-$41 \AA\ and He~{\sc ii} 4686 \AA\ lines produce strong and wide absorption artifacts, so the spectral window must be large enough for the QER20 Package to correct the artifacts without distorting the spectrum. This in turn  increases the demand for computational resources, which made it necessary to further optimize the algorithm. 

Figure~\ref{fig:reconstruidos} presents the reconstructed spectra for the stellar components of HDE~228766 returned by the QER20 Package, in the spectral range 4150 $-$ 4750 and in the H\,$\alpha$ and H\,$\beta$ lines. For the case of the He~{\sc ii} 4200 \AA, He~{\sc ii} 4542 \AA, N~{\sc v} 4604 \AA\ and H\,$\beta$ lines, we adjusted the algorithm in such a way that it only corrected the artifacts produced on the blue side of the main absorption. Thus, the emission feature of the P$-$Cygni profile of these lines is not mistaken for an artifact.

\cite{Massey1977} deduced approximately equal brightness for the components of HDE~228766 ($\Delta m_V \leq 0.5$ mag) by estimating the ratio of the intensities of the H8 and H9 spectral lines. From the analysis of the He~{\sc i} 4471 \AA\ and He~{\sc ii} 4542 \AA\ lines in observations corresponding to quadrature phases,~\cite{Rauw2002a} also concluded that both components are of similar brightness, with $m_{V,p} - m_{V,s} \approx 0.3$. These authors state that the overluminosity of the secondary could be due to the fact that the evolution of the whole system would have been influenced by mass exchange. We adopt this brightness ratio to estimate the contribution of each component to the total flux: 43.1\% for the primary and 56.9\% for the secondary. The reconstructed spectra presented in Figure~\ref{fig:reconstruidos} are rectified by these contributions.

Neither of the two reconstructed spectra exhibits the variable emission wings of the H\,$\gamma$ line that we identified in the composite spectra. This is because, during each iteration, the QER20 Package algorithm corrects the observations for the radial velocities of the stellar components, so any spectral features that do not follow this same pattern of velocities are diluted over the entire spectral window (this issue is discussed in Section~\S~\ref{sec:5_vientos}).

Our reconstructed spectra show that the H\,$\beta$ line of the primary is in absorption, while that of the secondary presents a clear P$-$Cygni profile. Meanwhile, the H\,$\alpha$ line appears in absorption for the primary and in strong emission for the secondary. The spectral line set N~{\sc iii} 4634$-$41 appears in emission in both components, very weak in the primary and strong in the secondary. The He~{\sc ii} 4200 \AA\ line appear in absorption in the spectra of both components, something typical of O stars, although this line presents an asymmetry in the secondary star and a weak P$-$Cygni profile in the primary. 

Contrary to~\cite{Hiltner1951} and in agreement with~\cite{Massey1977}, our spectral disentangling shows that the secondary of HDE~228766 has its own typical O$-$star absorption lines in the 4450 $-$ 4630 \AA\ window. These characteristic lines are He~{\sc i} 4471 \AA\ and He~{\sc ii} 4542 \AA.

~\cite{Rauw2002a} mention that in the observed spectra, the He~{\sc ii} 4542 \AA\ line presents a blue shift that could be the result of the strong wind of the secondary star. The spectral disentangling allows us to infer that the blue shift detected by~\cite{Rauw2002a} in this line is due to the weak asymmetric emission wings present in the reconstructed spectrum of the secondary, since the position of the maximum absorption remains at 4542 \AA. These emission wings are not artifacts, as they appear in the reconstructed spectrum of the secondary even though the QER20 Package applies the artifact correction strategy at each iteration. Moreover, the red wing is present in all observations (Figure~\ref{fig:reconstruidos}), while the blue wing appears in most of the observations. These emission wings in the secondary spectrum could be due to the superposition of photospheric absorption with the broad emission from rotating stellar wind material. 

\section{Evolutionary status}\label{sec:4_clasificacion}

\begin{figure*}
\centerline{\includegraphics[width=2.3\columnwidth]{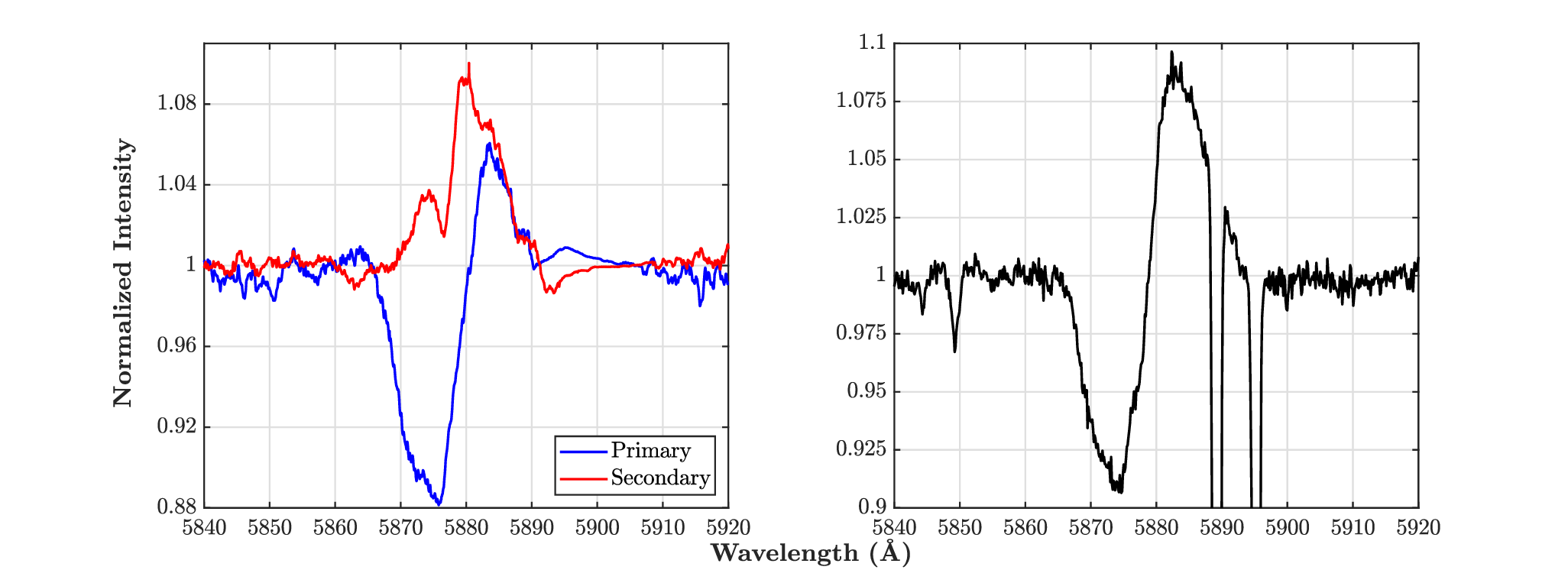}}
\caption{Left: Reconstructed spectra for the stellar components of HDE~228766 in the  He~{\sc i} 5876 \AA\ line, returned by QER20 Package. For disentangling we used only the 15 observations from the 2014 -- 2021 SPM campaigns (this line is not included in the spectral window of the OHP observations). The spectra are scaled assuming the light ratios 43.1\% and 56.9\% for the primary and secondary, respectively (see Section~\S~\ref{sec:3_separacion} for details).}Right: observation with the highest flux in the absorption feature of the P$-$Cygni profile. The intense and narrow absorptions at 5890 \AA\ and 5895 \AA\ correspond to the Na~{\sc i} of the interstellar medium (removed for disentangling).\label{fig:5876}
\end{figure*}

The reconstructed spectra of the stellar components of HDE~228766 allow us to perform a quantitative spectral classification of these stars for the first time. For that, we use the quantitative criteria of~\cite{Conti1971}, based on the flux ratio of the He~{\sc i} 4471 \AA\ and He~{\sc ii} 4542 \AA\ lines (see also~\cite{mathys1988final}, revisited by~\cite{Martins2018}).

For the primary star, $\log{\frac{EW(4471)}{EW(4542)}} = 0.09$, therefore this component is classified as O7.5. On the other hand, $EW(4686)$ $=$ 1.44 \AA, and according to~\cite{Martins2018} stars with $EW(4686)$ $>$ 0.6 \AA\ belong to luminosity class V with the qualifier ((f)). Furthermore,~\cite{Sota2011} and~\cite{Martins2018} state that the qualifier ((f)) is assigned in the case of weak N~{\sc iii} 4634$-$41 \AA\ emission and strong He~{\sc ii} 4686 \AA\ absorption, conditions that are fulfilled in the reconstructed spectrum of the primary component of HDE~228766. Finally,~\cite{Arias2016} state that the "z" characteristic is assigned when $\frac{EW(4686)}{EW(4471)} \geq 1.1$. For the primary component, $\frac{EW(4686)}{EW(4471)} = 2$. In conclusion, the primary star of HDE~228766 has the spectral classification O7.5 V((f))z.

The "z" characteristic was introduced by~\cite{Walborn1997} to identify O$-$type stars belonging to luminosity class V and exhibiting the He~{\sc ii} 4686 \AA\ line in strong absorption. Vz stars are the least luminous and, consequently, the closest to the zero$-$age main sequence (ZAMS) within the O$-$type stars. Although the quantitative studies of~\cite{Sabin-Sanjulian2014} reported numerous exceptions to this hypothesis,~\cite{Arias2016} found that almost all OVz stars from the Galactic O$-$Star Spectroscopic Survey (GOSSS) are associated with very young clusters, whereas clusters without Vz objects are older. These facts add weight to the interpretation of the "z" characteristic as a signature of youth.

Based on a qualitative analysis of the blended spectra,~\cite{Rauw2002a} suggest that the primary is the least evolved component of the binary HDE~228766. These authors explain this situation by a rejuvenation of the primary after a Roche lobe overflow (RLOF) of the secondary star. Our discovery of the "z" characteristic was made possible by our efficiency QER disentangling algorithm. Together with the mass ratio and the minimum masses yielded by the orbital solution (see Table~\ref{tab:losp}), it confirms the hypothesis of the mass transfer episode.
 
However the evolutionary status of the secondary is clearly the most salient question regarding HDE~228766. Is it a Wolf$-$Rayet star as suggested by~\cite{Rauw2002a}? This question bears on all other key aspects: the wind-wind interactions, the mass transfer and the rejuvenation of the primary. 

In the literature we find five criteria to differentiate between Of and late WN (WNL) spectral types:

\textit{\textbf{(i) Equivalent widths of the He~{\sc ii} 4686 \AA\ and He~{\sc i} 5876 \AA\ lines.}}

According to~\cite{Crowther1997}, Of stars are delimited by $\log{[EW(4686)]}$ $<$ $\approx$1  and $\log{[EW(5876)]}$ $<$ $\approx$0.6. Although the QER20 Package yields an adequate reconstruction of He~{\sc ii} 4686 \AA\ (see Figure~\ref{fig:reconstruidos}), unfortunately He~{\sc i} 5876 \AA\ is not included in the spectral window of the OHP observations. Therefore, we only have the 15 observations from the 2014 -- 2021 SPM campaigns to disentangle this line. Nevertheless, we perform the separation of these 15 spectra using the QER20 Package in order to obtain an approximation to the real aspect of the He~{\sc i} 5876 \AA\ line profile in the spectrum of the secondary, and an upper limit for the flux of this line. 

In the observed spectra of HDE~228766, the He~{\sc i} 5876 \AA\ line profile corresponds to the blending of the absorption of the primary star with the P$-$Cygni profile of the secondary. Therefore, we adjusted the QER20 Package algorithm so that the artifact correction is only performed on the blue side of the spectral line. Thus, we eliminate the risk that the emission feature of the P$-$Cygni profile is mistaken for an artifact. The left panel of Figure~\ref{fig:5876} presents the reconstructed spectra for the stellar components of HDE~228766 in the He~{\sc i} 5876 \AA\ line, returned by QER20 Package. For comparison, the right panel of Figure~\ref{fig:5876} shows the observation with the highest flux in the absorption feature of the P$-$Cygni profile. Before disentangling we remove the strong Na~{\sc i} interstellar lines at 5890 \AA\ and 5895 \AA.

The absorption feature in the reconstructed spectrum of the secondary and the strong P$-$Cygni profile in the primary are indications that disentangling is not effective in this line due to the low number of observations. Yet, this reconstruction is useful to estimate the upper limit of the flux of the He~{\sc i} 5876 \AA\ line in the secondary component: $EW(5876)_{max}$ $=$ $0.88$ \AA\ and $\log{[EW(5876)]}_{max}$ $=$ $-0.06$. Now, if we consider the extreme case in which the entire flux of the observation in the right panel of Figure~\ref{fig:5876} is attributed to the secondary component, we would obtain $EW(5876)_{max}$ $=$ $2.34$ \AA\ and $\log{[EW(5876)]}_{max}$ $=$ $0.37$. 

On the other hand, from the spectrum of the secondary component reconstructed by the QER20 package (Figure~\ref{fig:reconstruidos}), we obtain $EW(4686)$ $=$ $8.3$ \AA\ and $\log{[EW(4686)]}$ $=$ $0.92$. This result, and our maximum limit for $\log{[EW(5686)]}$, are below the maximum limits established in criterion \textit{(i)}, demonstrating that the secondary of HDE~228766 belongs to the spectral type Of.

\textit{\textbf{(ii) Maximum normalized intensities of the emission features of the P$-$Cygni profiles of the H\,$\beta$, H\,$\gamma$ and He~{\sc i} 5876 \AA\ lines.}}

In WNL stars, the emission features of the P$-$Cygni profiles of the H\,$\beta$, H\,$\gamma$ and He~{\sc i} 5876 \AA\ lines have much stronger maximum normalized intensities than mid$-$ or late$-$Of stars: $\approx +0.3$, $\approx +0.15$ and $\approx +0.5$, respectively~\citep{Crowther1997a}. In contrast, the maximum normalized intensities of these lines in the reconstructed spectrum of the secondary star of HDE~228766 are well below these values: H\,$\beta$ $=$ +0.07, H\,$\gamma$ $=$ 0.0 and He~{\sc i}$_{max}$ 5876 \AA\ $=$ +0.18.

\textit{\textbf{(iii) Maximum normalized intensity of the H\,$\alpha$ line.}}

Furthermore, WNL stars exhibit strong H\,$\alpha$ emission with maximum normalized intensities of $\approx +1.0$~\citep{Crowther1997a}, whereas the reconstructed spectrum of the secondary component of HDE~228766 shows a maximum normalized intensity of +0.67 for this line.

\textit{\textbf{(iv) Presence of P$-$Cygni profile in He~{\sc ii} 4542 \AA\, or blue shift if this line is only in absorption, in combination with $EW(5876)$ $\geqq$ 3 \AA\ and $EW(4686)$ $\geqq$ 12 \AA.}}

\cite{Bohannan1999}, mention that an Of star should be reclassified as WNL if its spectrum meets any of the following conditions: \textit{"1. If He~{\sc ii} 4542 \AA\ has a clear P$-$Cygni profile, or if it is only in absorption and has a clearly blue-shifted central wavelength of more than 50 km\,s$^{-1}$."}. In the secondary star of HDE~228766, the He~{\sc ii} 4542 \AA\ line does not have a P$-$Cygni profile but rather faint emission wings. Furthermore, although this line shows a slight asymmetry towards the blue, the wavelength at which the maximum absorption occurs remains at 4542 \AA. \textit{"2. If $EW(5876)$ $\geqq$ 3 \AA, with He~{\sc ii} 4686 \AA\ in emission."}. This condition is not fulfilled in the secondary star of HDE~228766, because although He~{\sc ii} 4686 \AA\ is in emission, $EW(5876)_{max}$ $=$ $2.34$ \AA. \textit{"3. If $EW(4686)$ $\geqq$ 12 \AA."}. This condition is not fulfilled in the secondary star of HDE~228766 since $EW(4686)$ $=$ $8.3$ \AA.

\textit{\textbf{(v) He~{\sc i} 4471 \AA\ line profile.}}

\cite{Crowther2011} recommend keeping the Of classification in mid$-$ and late$-$O stars if the He~{\sc i} 4471 \AA\ line is in absorption, which is the case of the secondary star of HDE~228766. Likewise,~\cite{Crowther1997a} and~\cite{,Crowther2011} indicate that the stars WN8$-$9 present a very strong P$-$Cygni profile in the He~{\sc i} 4471 \AA\ line, while the mid$-$ or late$-$Of stars show this line in absorption, as happens with the secondary component of HDE~228766.

Since the application of the five criteria on the reconstructed spectrum of the secondary component of HDE~228766 unanimously confirms that this star has the Of spectral type and not the WNL, we can now perform its quantitative spectral classification. According to~\cite{Conti1971} and~\cite{Martins2018}, 
the ratio $\log{\frac{EW(4471)}{EW(4542)}} = -0.26$ yields an O6 type. Since $EW(4686)$ $=$ $-8.3$ \AA, the secondary component of HDE~228766 belongs to luminosity class Ia with the qualifier f. Furthermore,~\cite{Sota2011} state that the qualifier f is assigned when the N~{\sc iii} 4634-41-42 \AA\ and He~{\sc ii} 4686 \AA\ lines are in emission above the continuum. In conclusion, the secondary component of HDE~228766 is O6 Iaf.

Even though our spectral analysis places the secondary component of HDE~228766 at a pre$-$Wolf$-$Rayet evolutionary stage, several features of the reconstructed spectrum clearly indicate that this star is already in transition to the WN stage, namely the emission features of the P$-$Cygni profiles of the H\,$\beta$ and He~{\sc i} 5876 \AA\ lines.

\section{H\,$\gamma$ variable emission}\label{sec:5_vientos}

\begin{figure}
    \centerline{\includegraphics[width=\columnwidth]{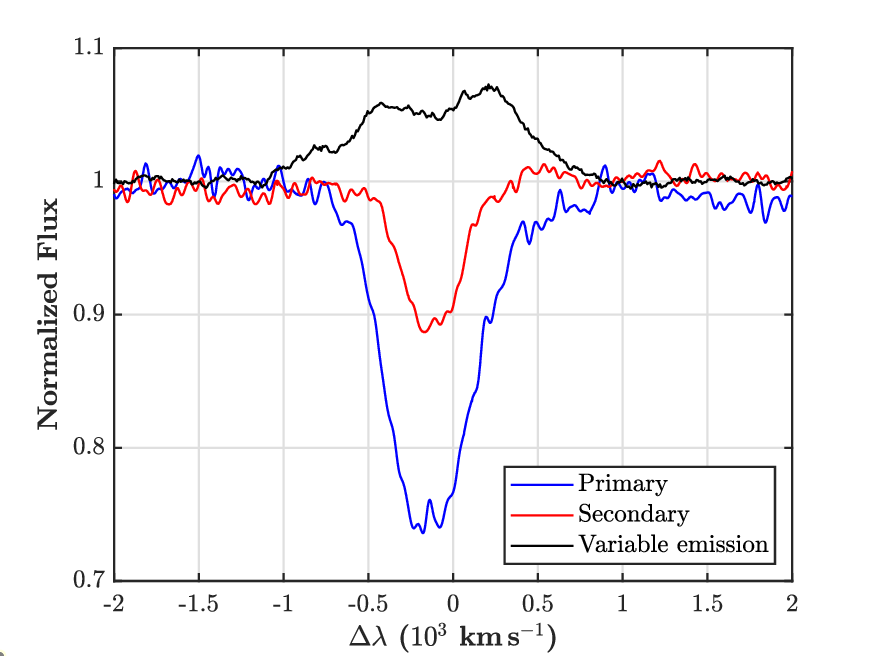}}
	\caption{Reconstructed spectra for the stellar components of HDE~228766 in the H\,$\gamma$ line, using only the 20 OHP observations from the 1999 campaign. The spectra are scaled assuming the light ratios 43.1\% and 56.9\% for the primary and secondary, respectively. Black: Average of the 15 spectra obtained by subtracting the composite spectra that reproduce the 1999 observations from the 15 observations of the 2014 to 2021 campaigns, for the same phases of the orbital cycle (see text for details).\label{fig:HgammaQERSolo1999}}
\end{figure}

As mentioned in \S~\ref{subsec:2.1_spec} a variable emission feature is present at H\,$\gamma$ in all our 11 spectra from the 2014 campaign, whilst it is not seen in the 1999 spectra  
(Figure~\ref{fig:1999vs2014}). This feature was also observed in each of our three 2015 spectra, as well as in our 2021 spectrum.

In WR + O binaries, the variable emission associated with their colliding winds 
is generally observed as an excess emission superposed to the strong WR emission lines. 
This typically occurs in the He~{\sc ii} 4686 line, as is the case with the WN + O systems WR 21, WR 31, WR 47~\citep{Fahed2012} and WR 139~\citep{Flores2001}; or in the C~{\sc iii} 5696 and C~{\sc iv} 5808 lines, as is the case with the WC/WO + O binaries WR 9, WR 30a~\citep{Bartzakos2001}, WR 42~\citep{Hill2000}, WR 48~\citep{Hill2002} and WR 79~\citep{Luehrs1997}.

\begin{figure*}
\centerline{\includegraphics[width=2\columnwidth]{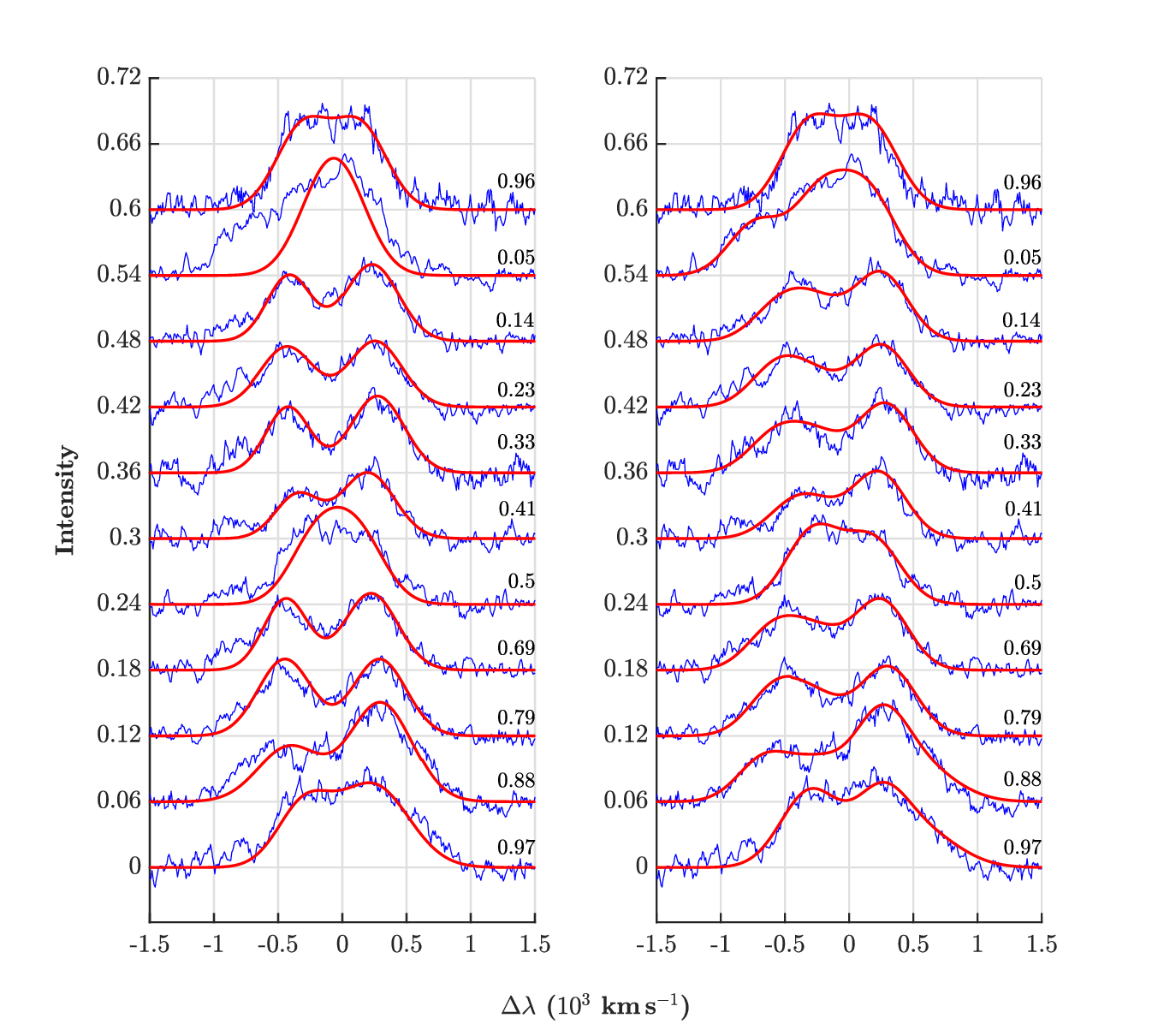}}
\caption{Left: Fitting with two Gaussians (red) of the variable emission feature present in the H\,$\gamma$ line of the 2014 observations (blue). Right: The same as in the left panel, only performing the fit with four Gaussian curves (two fixed at -240 and +240 km\,s$^{-1}$, and two moving).The orbital phase is indicated to the right of each panel.\label{fig:Obs2014_2Gauss_Vs_4Gauss}}
\end{figure*}

In the case of HDE~228766 (classified in this work as O7.5 V + O6 Iaf), the H\,$\gamma$ line is in absorption both in the composite spectra and in the reconstructed spectrum of the individual components (Figure~\ref{fig:reconstruidos}). This makes it difficult to analyze the variable emission feature, as it is blended with the main absorption.

To overcome this difficulty and allow a detailed analysis of the variable emission observed since 2014 in the H\,$\gamma$ line, we need to subtract the absorption feature. This can be done because the 1999 spectra are in pure absorption.

\begin{figure*}
\centerline{\includegraphics[width=2.1\columnwidth]{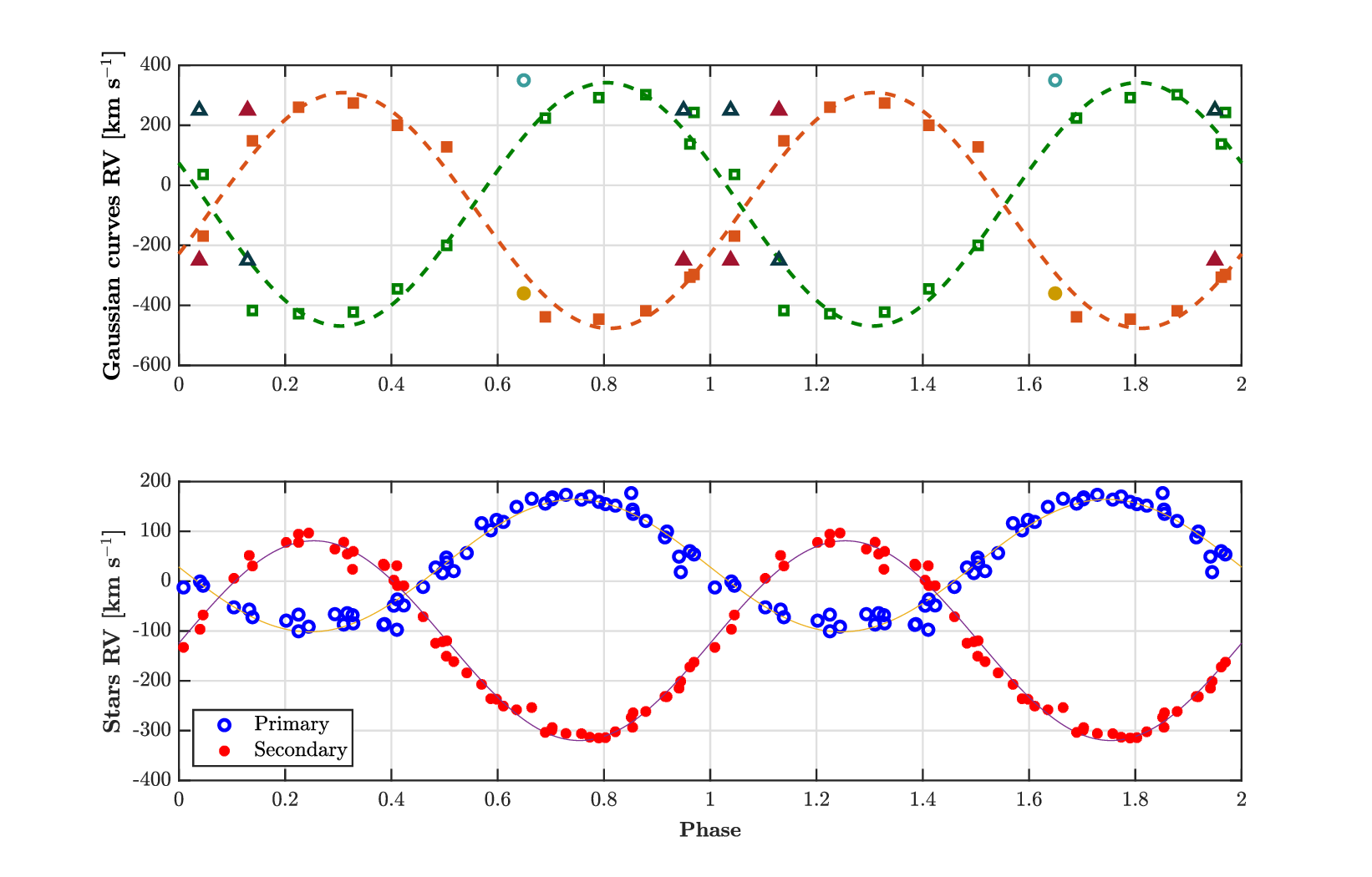}}
\caption{Top: Sinusoidal fits to the radial velocities of the two Gaussians we used to fit the variable emission feature of the H\,$\gamma$ line present in the observations from 2014 (Figure~\ref{fig:Obs2014_2Gauss_Vs_4Gauss}, left) to 2021 (Figure~\ref{fig:Obs2015_2021}). 2014: green open and orange filled squares. 2015: blue open and brown filled triangles. 2021: cyan open and yellow filled circles. Below: Radial velocities of the primary (blue) and secondary (red) components of HDE~228766 (see Figure~\ref{fig:RVs}).\label{fig:RVs_Dos_Gauss}}
\end{figure*}

To reduce noise and to perform the subtraction at the exact orbital phases of the 2014 -- 2021 observations, we apply again our QER20 disentangling algorithm around the H\,$\gamma$ line, but this time using only the 20 observations of 1999. Having thus obtained the averaged absorption spectrum of each stellar component (red and blue spectra in Figure~\ref{fig:HgammaQERSolo1999}), these can be subtracted, after the proper redshift correction, from each of the 15 observations of the 2014 -- 2021 campaigns.

Figure~\ref{fig:HgammaQERSolo1999} shows in black the average spectrum after subtraction. Note that the average of the variable emission feature ranges from $\approx$$-970$ to $\approx$970 km\,s$^{-1}$, with a slight decrease in intensity at the central wavelength. In addition, the emission flux is considerable compared to the absorption flux (presumably from the stellar components). This is evidence that HDE~228766 is a particular case, since in most of the binaries showing wind$-$wind collisions reported in the literature, the excess emission originating in the interaction between the winds has a flux much lower than the stellar emission. 

In order to establish the possible origin of this emission, in a first approach, we fit the 15 emission   2014 -- 2021 spectra with two Gaussian curves, fixing their width at 900 km\,s$^{-1}$, allowing their flux to vary up to $\pm$10$\%$, and leaving the radial velocity associated with the Doppler shift as a free parameter. The left panel of Figure~\ref{fig:Obs2014_2Gauss_Vs_4Gauss} shows that a rather good fit is obtained for most phases observed in 2014. Yet it is clear that two Gaussians are not capable to reproduce the observations at 
phases near conjunction (notably phases 0.05 and 0.5). 

\begin{figure}
	\centerline{\includegraphics[width=\columnwidth]{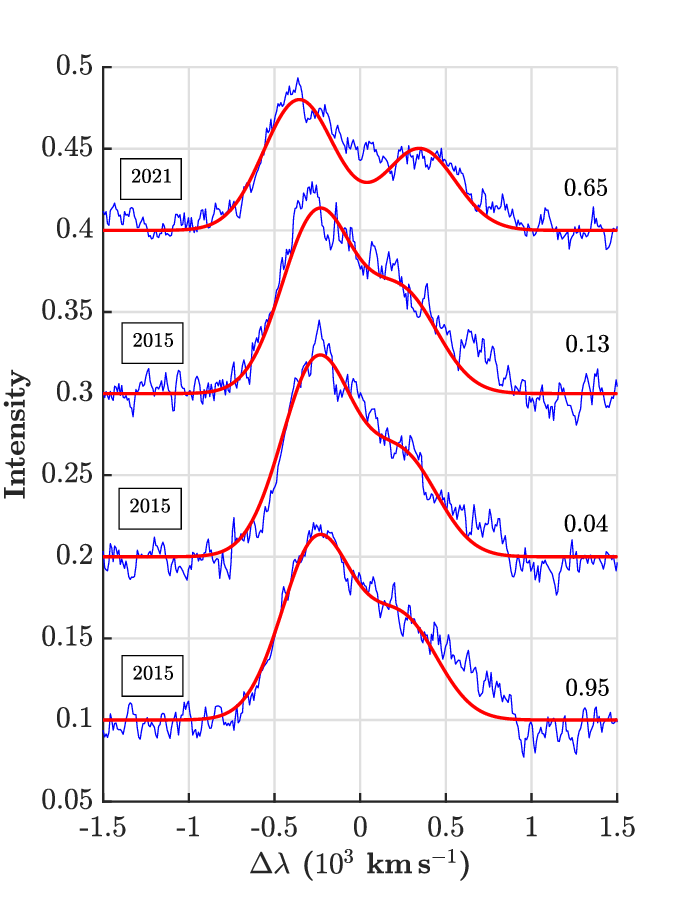}}
	\caption{Variable emission feature in the H\,$\gamma$ line present in the observations from 2015 to 2021 (blue), with their respective fitting curves using two Gaussians (red). In the case of the three 2015 observations, the best fit is obtained when the two Gaussians have fixed velocities centered at +240 and -240 km\,s$^{-1}$, despite the fact that these are observations at different phases of the orbital cycle (phases are indicated on the right).\label{fig:Obs2015_2021}}
\end{figure}

If the gas represented by these two Gaussians follows the orbital motion, its radial velocity must show a sinusoidal curve because the orbit is circular. This is indeed what we find for the 2014 spectra (Figure~\ref{fig:RVs_Dos_Gauss}). However neither the 2015 observations at phases 0.04 and 0.13 nor the 2021 observation follow this sinusoidal  curve. On the contrary, we find that the best fit for each of the three 2015 spectra is obtained with two Gaussians centered at +240 and -240 km\,s$^{-1}$. This seem to indicate that they are produced by gas maintaining a constant Doppler shift, despite the fact that these observations correspond to different phases of the orbital cycle (Figure~\ref{fig:Obs2015_2021}).

Thus our simple model of two Gaussians following the orbital motion breaks down because of (1) the inadequate fit of the line profile at phases near conjunctions and (2) the fact that the three 2015 observations are best fitted with two Gaussians having fixed radial velocities. This lead us to infer that, superposed on the two emissions produced by the wind$-$wind collision (which follow the orbital motion), there is a stationary emission coming from a spatially extended zone. 

Consequently, we proceed to repeat the fitting process but now using four Gaussians: two with fixed radial velocities centered around -240 and +240 km\,s$^{-1}$, and two having free velocities. Again we set the width at 900 km\,s$^{-1}$ and allow a maximum variation of $\pm$10$\%$ in the flux. This implies that the gas producing the emission would occupy a region that spatially spans beyond the orbits of the stars, since in velocity space the primary and secondary components have velocities of 134 and 200.5 km\,s$^{-1}$, respectively (see Table~\ref{tab:losp}). The right panel of Figure~\ref{fig:Obs2014_2Gauss_Vs_4Gauss} shows that the new fit with four Gaussians  reproduces much better the observations. This seems to indicate that there are four components producing the H\,$\gamma$ emission: two are stationary and two are moving. 

\begin{figure}
	\centerline{\includegraphics[width=0.95\columnwidth]{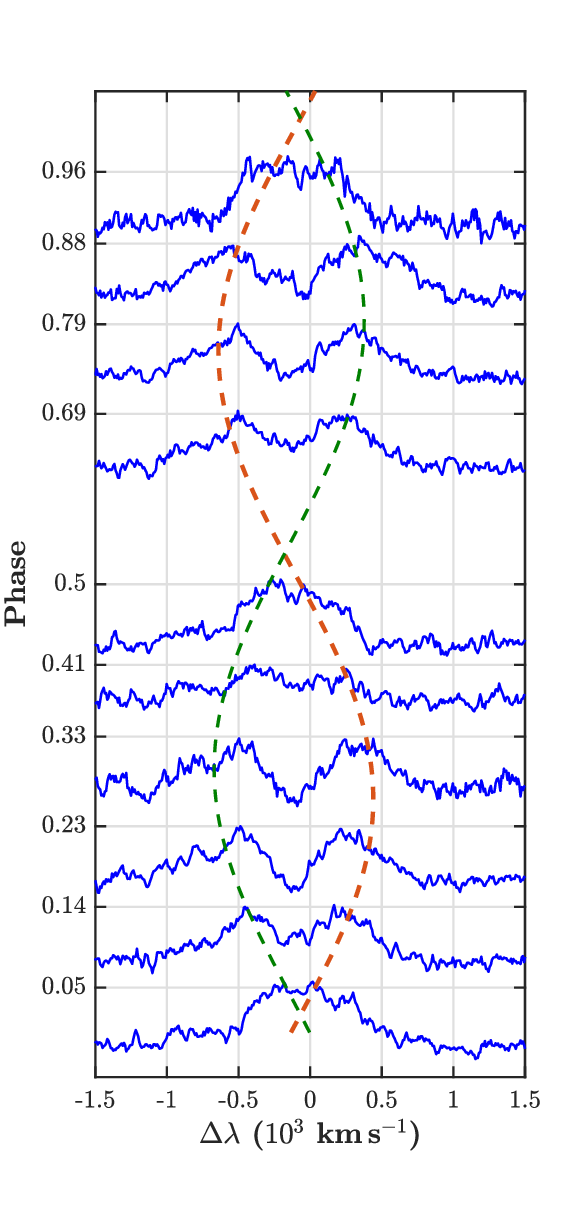}}
	\caption{Variable emission feature present in the H\,$\gamma$ line of the 2014 observations, after subtraction of the two emissions with fixed Doppler shifts (-240 and +240 km\,s$^{-1}$). The green and orange curves correspond to the fit curves illustrated in the upper panel of Figure~\ref{fig:RVsStarsAndShockCone}.\label{fig:Ajuste10Obs2014SinGaussFijas}}
\end{figure}

\begin{figure*}
\centerline{\includegraphics[width=2.1\columnwidth]{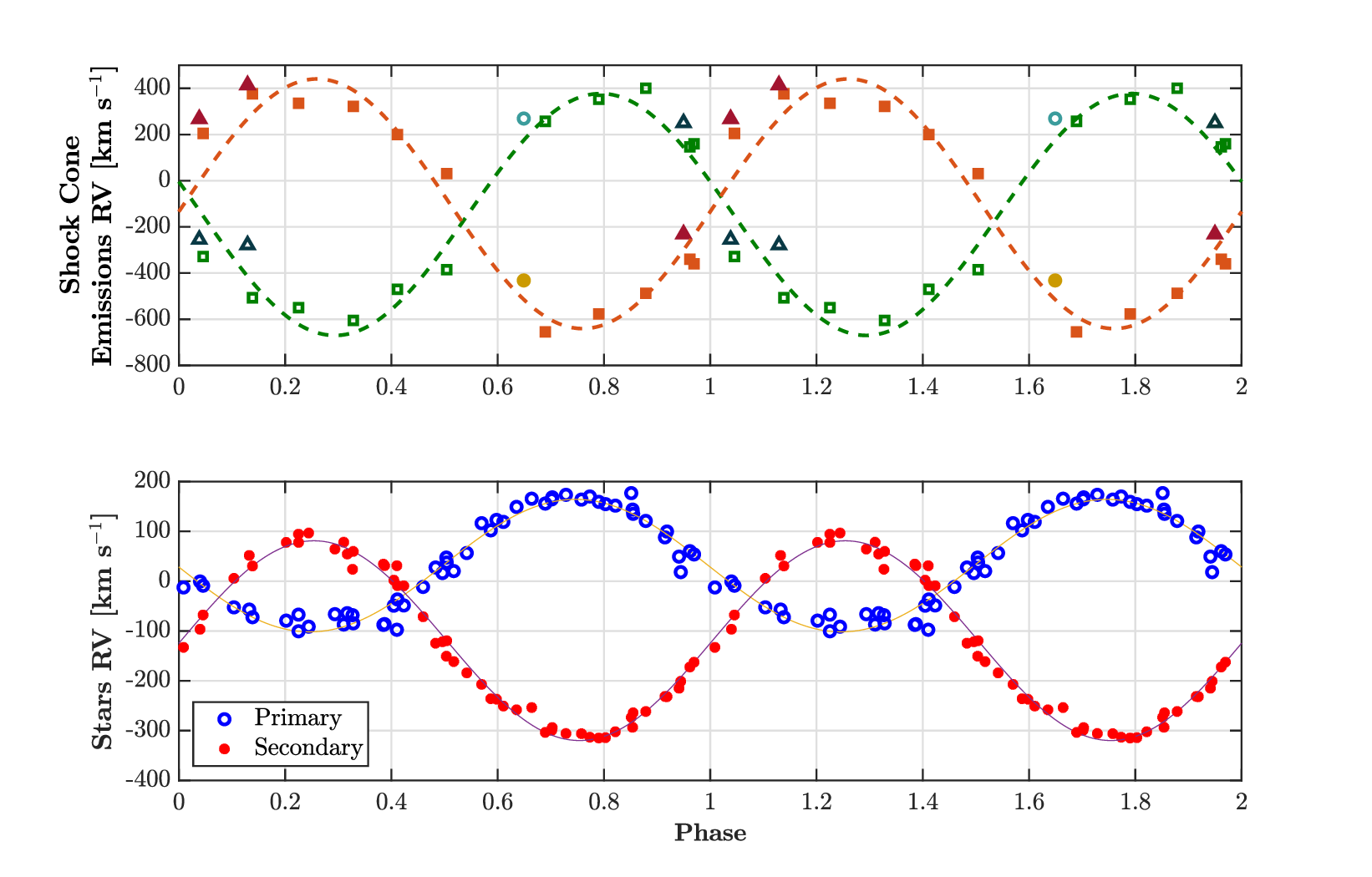}}
\caption{Top: Sinusoidal fits to the radial velocities of the variable emission feature present in the H\,$\gamma$ line of the 2014 -- 2021 observations, after subtraction of the two emissions with fixed Doppler shifts (-240 and +240 km\,s$^{-1}$). 2014: green open and orange filled squares. 2015: blue open and brown filled triangles. 2021: cyan open and yellow filled circles. Bottom: Radial velocities of the primary (blue) and secondary (red) components of HDE~228766 (see Figure~\ref{fig:RVs}).\label{fig:RVsStarsAndShockCone}}
\end{figure*}

Figure~\ref{fig:Ajuste10Obs2014SinGaussFijas} presents the result obtained by subtracting the two fixed radial velocity Gaussians from the variable emission feature present in the 2014 observations. After removing the two stationary emissions, two emissions with Doppler shifts that change in phase with the orbital motion are clearly identified (Figure~\ref{fig:RVsStarsAndShockCone}). 


Our analysis suggests that the emission zones are spatially  extended around the stellar components, which supports our hypothesis that the wind shock cone is insufficient to explain the excess emission we discovered in the observations of the 2014 to 2021 campaigns. In fact, this region extends far beyond the orbit of the secondary. This explains why the average of the variable emission feature has a width ($\pm$ $\approx$$970$ km\,s$^{-1}$, Figure~\ref{fig:HgammaQERSolo1999}) greater than the semi$-$amplitude of the radial velocity curve of this component ($\approx$200.5 km\,s$^{-1}$, Table~\ref{tab:losp}). 

What could be the origin of these emission zones? WR $+$ O star binaries have a highly non$-$conservative evolution~\citep{petrovic2020}. In this scenario, when mass is transferred from the donor to the mass$-$gainer, either not all of the transferred matter ends up on the mass gainer or the mass gainer immediately ejects some of it~\citep{Langer2022}. Our hypothesis is that the binary HDE~228766 undergoes a non$-$conservative evolution. Thus, a Roche lobe overflow episode produced the rejuvenation of the primary component and led to the current mass ratio and minimum masses. During the mass exchange, part of the material ended up distributed in a spatially extended region around the stellar components, producing the excess emission that we discovered in the H\,$\gamma$ line.

\section{Conclusions}\label{sec:conclusions}

Using new observations of the massive binary HDE~228766, we update the orbital parameters of this system. We report for the first time a variable emission feature in the H\,$\gamma$ line, present during our 2014 to 2021 observing campaigns.  
We apply to this binary our novel QER20 disentangling Package, yielding the first reconstructed spectra of the stellar components.  These enable us to perform a quantitative spectral classification of the components for the first time: O7.5 V((f))z for the primary component and O6 Iaf for the secondary. The "z" characteristic in the classification of the primary star indicates that the secondary is the most evolved component, which supports the hypothesis that the whole system could have been influenced by mass exchange. This hypothesis is consistent with the high mass ratio of the system, the elevated minimum masses of the stellar components and the overluminosity of the secondary. In addition, the reconstructed spectrum of the secondary star contains clear indications that it is in an early stage of transition to the WN state: relatively weak emission features of the P$-$Cygni profiles of the H\,$\beta$ and He~{\sc i} 5876 \AA\ lines. 

We also conclude that the variable H\,$\gamma$ emission is composed of at least four independent features. Two of them are stationary and the other two follow the orbital motion. Thanks to the QER20 Package we are able to isolate this emission for analysis. This reveals the existence of a spatially extended emission zone that has not been reported so far. This zone could be the result of an episode of enhanced mass loss in a non$-$conservative evolution scenario; either from the secondary toward the primary or from the whole system.

We have started an observing program to monitor the medium$-$term behavior of the variability of the H\,$\gamma$ line, and possibly to detect variability in other lines. This will enable a more detailed study of the region(s) where it originates. Modeling of these new spectra using a stellar atmosphere code (e.g. CMFGEN), and additional X$-$ray observations, will enable us to describe the shock zone and better estimate the physical parameters of the wind and the stellar components. New observations in the optical should allow us to reliably reconstruct the profile of the He~{\sc i} 5876 \AA\ line in the secondary star and help us monitor its transition toward the Wolf$-$Rayet WNL stage.

\section*{Acknowledgments}

We are grateful to Prof. Hugues Sana for the LOSP Package, and to Prof. Gregor Rauw for comments on an earlier version of this paper and for sharing with us his spectra of HDE 228766 collected between 1999 and 2021 at the Observatoire de Haute Provence (OHP), France. We acknowledge the financial support of the Consejo Nacional de Humanidades, Ciencias y Tecnolog\'ias de M\'exico (CONACyT) and the Vicerrector\'ia de Investigaciones Innovaci\'on y Extensi\'on of the Universidad Tecnol\'ogica de Pereira (Colombia). We thank the referee for constructive comments that have helped to improve the paper. We also thank the staff of San Pedro M\'artir Observatory (M\'exico).This research has made use of NASA's Astrophysics Data System Bibliographic Services.

\section*{Data Availability}

The SPM spectra can be made
available upon reasonable request.



\bibliographystyle{mnras}
\bibliography{Disentangling_Biblio} 



\bsp	
\label{lastpage}
\end{document}